# "SHARING WISDOMS FROM THE EAST": DEVELOPING A NATIVE THEORY OF ICT4D USING GROUNDED THEORY METHODOLOGY (GTM) – EXPERIENCE FROM TIMOR-LESTE

Abel Pires da Silva, Centro de Estudos Estratégicos de Timor-Leste (CESTIL), abel.dasilva@alumni.anu.edu.au

**Abstract:** There have been repeated calls made for theory-building studies in ICT4D research to solidify the existence of this research field. However, theory-building studies are not yet common, even though ICT4D as a research domain is a promising venue to develop *native* and *indigenous* theories. To this end, this paper outlines a theory-building study in ICT4D, based on the author's experience in developing a mid-range theory called 'Cultivating-Sustainability' of E-government projects, a native mid-range theory of ICT4D. The paper synthesizes the GTM literature and provides a step-by-step illustration of GTM use in practice for research students and early career ICT4D academics. It introduces the key strategies and principles of GTM, such as the theoretical sampling strategy, the constant comparison strategy, the concept-emergent principle, and the use of literature throughout the study process. Then discusses the steps involved in the data collection and analysis process to develop a theory using case studies as sources of empirical data; it concludes with a discussion on using the strategies and principles in the three case studies. It is expected that this paper contributes to the diversification of research methodology, particularly to our collective quest for developing native and indigenous theories in the ICT4D research domain.

## 1. INTRODUCTION

Theory development studies in ICT4D have long been encouraged to legitimize and recognize the existence of ICT4D as a separate academic discipline (Sahay & Walsham, 1995). However, progress has been slow on this front (Avgerou, 2017; Walsham, 2017). It has been noted that most ICT4D studies are focusing on the description of findings and the use of theory to help explain their findings (Lin et al., 2015). Few published studies have attempted to develop any forms of theory based on empirical evidence extracted from ICT4D initiatives/projects.

Even though GTM's use in the information systems (IS) field research has proliferated in the last decades (Wiesche et al., 2017), its application in the ICT4D area remains low. Even fewer are the guidelines for theory-building studies in the ICT4D context. This paper aims at promoting theory development studies in ICT4D by showcasing a process to develop a substantive theory called "Cultivating Sustainability" of E-gov projects, using GTM and the case study method. The theory itself has been discussed thoroughly by Da Silva & Fernández (2020), and this paper focuses on the methodological aspects of the theory-building process. According to Straub (2012, p. v), "a native (indigenous) theory is a theory specifically developed to describe, explain, predict, or design IS [or ICT4D] phenomena". As this theory-building study was based on three E-government projects in Timor-Leste, a developing country, the theory is a *native* theory of ICT4D.

The paper contributes in two ways: (1) synthesize the GTM literature (using data collected from case studies) for research students and early career academics, and (2) provide a step-by-step illustration of GTM use in practice. It is expected that the theory development process outlined in this paper may be used as a reference for future studies to develop more native theories of ICT4D and also to identify and construct *indigenous* theories in ICT4D based on the local wisdom of the societies living in the less-developed countries (LDCs).

This paper *first* presents an overview of GTM, its strategies, and principles, its data collection and analysis process, *then* outlines the use of GTM to develop a substantive theory in ICT4D using data from three (3) case studies of E-government projects; and *lastly*, the conclusion.

## 2.   GROUNDED THEORY METHODOLOGY (GTM)

GTM was first introduced more than 50 years ago by Barney Glaser and Anselm Strauss in 1967 and is now one of the most widely used qualitative research methodologies in IS research (Wiesche et al., 2017) and also globally (Birks & Mills, 2011; Gummesson et al., 2007). Out of the main three variants of GTM available, the Glaserian (classic) GTM was adopted in this paper simply because of its close adherence to the original teaching of the GTM introduced by Glaser and Strauss (Glaser & Strauss, 1967).

In GTM, theories are developed inductively from data through the incremental and systematic progression of knowledge based on real-life cases (Glaser & Strauss, 1967). Thus, the theory will be able to explain the subject studied, fit, and relevant to that particular research topic. The inductive process of exploration also facilitates our understanding of complex organizational phenomena (Locke, 2001; Martin & Turner, 1986; Ven & Poole, 1989), such as the complex change process that takes place when delivering ICT4D technologies in LDCs (Avgerou, 2001). Also, the theory is modifiable as new data are collected (Glaser, 1998). The inductive theory development process is the main differentiating factor between GTM and other qualitative research methods (Urquhart et al., 2010).

## 3.   GTM AND CASE STUDY

In GTM, data collection and analysis are conducted simultaneously (Urquhart et al., 2010); however, GTM only specifies the analytic strategy and not the methods for data collection (Charmaz, 2003). For this reason, this study follows Walsham's (1995) discussion on the use of GTM in interpretive case study research to collect primary data.

The integration of a case study in GTM and its validity and reliability has been thoroughly discussed by Diaz-Andrade (2009); therefore, this paper focuses on the mechanization of the theory-building process. The use of case studies is essential for IS/ICT4D researches because it brings the focus of study on 'contemporary phenomenon within a real-life context' (Yin, 2009, p. 2); as such, it is 'well-suited to capturing the knowledge of practitioners' (Benbasat et al., 1987, p. 370). The combination of GTM and case studies allows researchers to link their research with the latest developments in practice (Benbasat et al., 1987). Thus, it enables the achievement of dual objectives of rigor and relevance because the theories developed are firmly grounded to empirical reality (Benbasat & Zmud, 1999; Fernández & Lehmann, 2005; Glaser & Strauss, 1967).

In conducting a GTM study using case studies, it is crucial to begin the study with a single case called the foundation case that serves as the gateway to the theory development study in the proposed research topic (Lehmann, 2001). Subsequent cases will be selected based on the results of the first case.

## 4.   GTM DATA ANALYSIS PROCEDURE

In GTM, the analytic strategy can be best presented as a package that guides a researcher from the first moment entering the field (where the researcher still has very little knowledge of the research subject) to the final publication (where the researchers have become theorists in their respective areas); the package detailed steps is known as 'double-back steps' (Glaser, 1998).

The double-back steps are described as moving back and forth throughout the research activity (Glaser, 1978, p. 16) and include data collection, open coding, theoretical sampling, memo writings to facilitate the coding and sampling process until the emergence of the core concepts. The core concepts then guide more selective coding, sampling strategies, and memoing. These double-back steps continue iteratively until the findings are saturated, at which point the analyst continues to the sorting of memos to produce theoretical frameworks. After the sorting step, the analyst starts working on the publication of the research findings.

Throughout the research process, GTM researchers are guided by the following two strategies and one principle: (1) 'theoretical sampling,' (2) 'constant comparison' strategies, and (3) the 'conceptual emergent' principle. These three characteristics are facilitated by the use of 'memo writing' and engagement with the literature. The following sub-sections discuss these characteristics of GTM.

### 4.1. Theoretical Sampling Strategy

The theoretical sampling strategy is the backbone of the GTM process to generate theory where data collection, coding, and analyses occur simultaneously. When using case studies, there are two main sampling strategies used in GTM: the 'intra-case sampling strategy' and the 'inter-case sampling strategy' (Fernández & Lehmann, 2011). The intra-case strategy focuses on 'selecting more "slices of data" from within each case to saturate concepts and maximize their conceptual yield' (Fernández & Lehmann, 2011, p. 9). Once no new findings can be obtained from one particular case, the next stage of sampling is to include more data from other cases (inter-case sampling). The selection of the new cases is guided by the findings that emerged so far from the existing case(s). Data collected from this inter-case sampling strategy will help to confirm and/or revise emergent (core) concepts and their properties, thus, help to saturate the findings (Fernández & Lehmann, 2011; Glaser & Strauss, 1967).

In order to develop theory, the data is analyzed in two phases (Christiansen, 2007; Glaser, 1978): 'substantive coding', which takes place during both the intra-case and inter-case sampling, and 'theoretical coding', which mainly takes place during inter-case sampling. Substantive coding is the coding in the substantive area with the primary aim to produce 'core concepts' in the area being studied. Two coding processes take place at this stage: *first*, the 'open coding' and *then* the 'selective coding' process. Open coding is the first coding process used to openly code data until the core concepts have emerged from data. This open coding process involves a rigorous investigation of data to generate concepts based on a set of empirical indicators by constantly comparing the data (Glaser, 1978, p. 62). The comparison continues: (1) between indicators, (2) between indicators and concepts, and (3) between concepts until the core concepts emerged (Glaser, 1978). These indicators may point to the existence of patterns of behavior, and the concepts can thus, be generated by naming these patterns.

Once the core concepts have emerged, the next step is to selectively code data only around the emerged core and its related sub-core concepts (selective coding process) until these concepts have been saturated. If there is more than one core concept emerged, the researcher has the freedom to select the most suitable concept for further investigation. Once the saturation has been reached, the next step is to focus on developing relationships between the core concept and its properties (sub-core concepts and their properties). This step is the 'theoretical coding' phase in the theoretical sampling of GTM (Glaser, 2005).

### 4.2. Constant Comparison

As discussed in the last sub-section, the constant comparison strategy is central to the GTM analysis process because it is 'the driving technique of GTM's data analysis [and] the facilitator of theoretical sampling' strategy (Urquhart and Fernández 2013, p.225). Glaser (1978) sets three types of comparisons in GTM to develop the theory: *first*, indicator to indicator to establish underlining uniformity and its varying conditions, both uniformity and the conditions are in the form of concepts; *second*, comparison between emerged concepts and more indicators, thus, generating new theoretical properties of the concepts; *third*, between the emerged concepts to produce an even higher degree of abstract concepts. This higher degree of conceptualization and the integration into the relationships (hypotheses) between concepts will produce a theory.

### 4.3. Conceptual Emergence

The double-back steps in the GTM research process discussed above will allow the full emergence of concepts that are truly developed from data. It is important to note that conceptualizations from data require the ability to *'lift'* data to an abstract level to understand 'what is going on in data' beyond a mere mechanistic effort to assign names into indicators (Glaser, 2002; Suddaby, 2006; Urquhart & Fernández, 2013). Conceptual emergence is an essential principle in GTM because, with it, the theory will explain the subject being studied, fit, and relevant to the research area. The theory will also be modifiable as new data becomes available (Glaser, 1998).

### 4.4. Memo Writing

Glaser (1978) emphasizes that memo writing in GTM studies is essential; it sits at the core of the theory generation process because 'memos are the theorizing write up of ideas about codes and their relationships as they strike the analyst while coding' (Glaser, 1978, p. 83). Memos also represent a researcher's thinking process about 'what is really happening in data,' and the literature consulted so far and how they relate to each other. Thus, memos are constantly revised as new understandings of data are developed from new slices of data.

Birks and Mills (2011) suggests useful types of memos for conducting GTM studies, and these are 'operational memos', 'coding memos' and 'analytical memos': (1) operational memos are written to record and facilitate the researcher's thinking about data collection and analysis process; (2) coding memos are written to explain what a particular code is, and the reasons behind the labeling of the code; and (3) analytical memos are written to record insights into conceptualizations developed from data and the relationships between concepts which will be integrated into the theory to facilitate 'theoretical coding' process.

### 4.5. Theoretical Sorting

The theoretical sorting step in GTM is conducted at the end of the study when the researcher is well into the fieldwork stage and has almost reached saturation (Glaser, 1978, pp. 116–127). Guided by the key question of 'where does [this memo] fit?', theoretical sorting of memo funds (that was developed throughout data collection and the analysis process) is a crucial step to formulate the emergent theory, and thus, it involves the sorting of concepts instead of data (Glaser, 1978, p. 116). This process involves a constant comparison strategy by moving back and forth between memos and the potential outlines of the theory, which in turn, produces more memos to be sorted. This process involves the promotion/demotion of concepts and facilitates the development of a core concept. This core concept might also be in the form of a process called the basic social process (BSP) in GTM's term (Glaser, 1978; Glaser & Strauss, 1967).

### 4.6. Core-Concept and BSP in GTM

GTM studies focus on identifying the core concept of the studied phenomena (or core 'category' in GTM's term). It is a concept that 'sums up the pattern of behavior', and thus, it is 'the substance of what is going on in the data' (Glaser, 1978, p. 94). A core category (concept) is the highest level of a category (concept) because it "relates to most other categories and their properties, [and] through these relations the core category accounts for most of the on-going behavior in the substantive area being researched" (Glaser, 1998, p. 135).

Due to this explanatory power, the emergence of the core concept will help researchers focus and sort the theory around it and explain the phenomenon studied using as few concepts as possible (Glaser, 1978). One type of core category/concept is BSPs. This type of concept has a process-out explanation power and has at least 'two or more clear emergent stages' (Glaser, 1978, p. 97). BSPs also have core properties of 'pervasiveness' and 'full variability' (Glaser & Holton, 2005). BSPs are pervasive because the concepts are fundamental, patterned processes in social organization, and thus 'unavoidable irrespective of conditional variations' (Glaser & Holton, 2005, p. 9). BSPs are also 'fully variable' because the concepts are abstract conceptualizations of data, which transcends beyond their original unit of analysis, and consequently, are stable and can account for change (Glaser and Holton 2005, p.10). Therefore, BSPs have the conceptual grasp and 'transcends, organizes and synthesizes large numbers of existing studies' (Glaser, 1992, p. 34). This power of BSPs becomes the main contribution of GTM studies because BSPs fulfill two prime attributes of theory: parsimony and scope (Glaser, 1992, p. 34).

### 4.7. Engaging Literature in GTM

In GTM, both 'theoretical sensitivity' and 'keeping an open mind' are essential parts of the methodology. Glaser (1992) states that theoretical sensitivity refers to researchers' knowledge, understanding, and skill to assist the generation of concepts from data. At the beginning of the study, theoretical sensitivity can be achieved through reading the literature (Martin, 2006; Urquhart & Fernández, 2013). However, researchers must keep their minds open when entering the field so that the sensitivity mentioned above will not contaminate their efforts to generate concepts from the data (Glaser, 1978, 2012).

Originally, as discussed by Glaser and Strauss (1967), there was a concern that the results of any literature review might potentially shape researchers' 'a priori' conceptualizations and could cause bias in dealing with data collection and analysis processes. Consequently, the condition may prevent researchers from discovering the actual truth from data. However, there has been a shift away from this concern when Glaser (1978, p. 45) suggested that researchers can start with a general perspective with beginning concepts and field research strategies. The underlying argument is that it is possible for a researcher to access the literature without accepting it as the final truth (Walsham, 1995).

To facilitate the use of literature in GTM studies, Martin (2006) outlines four steps in engaging with the literature: (1) noncommittal, (2) comparative, (3) integrative, and (4) transcendent phase. These steps are discussed as follows:

### 4.7.1. Noncommittal

Glaser (1998) acknowledges the situation where avoiding literature is impossible, especially when the researcher is trying to fulfill PhD formal requirements or when the researcher is applying for a research grant. Therefore, Glaser (1998) suggests that the review of literature is conducted with the attitude of data collection, thus, treating the literature as data to be compared with other data obtained from the field in the later stages of the research. The review is conducted 'on the fundamental understanding that the generated grounded theory will determine the relevance of the literature' (Urquhart & Fernández, 2013, p. 230).

This noncommittal step helps the researchers avoid being committed to specific theories at the beginning of the research and the subsequent abandonment of the theories at a later stage when they became irrelevant to the emergent findings. For instance, Walsham and Sahay (1999) shared their experience where initially their study was influenced by structuration theory and social construction of technology. Later on, the study moved to focus on the actor-network theory (ANT); they shared that (p. 41):

"[T]he theoretical basis of the study evolved over time in response to both our deepening understanding gained through the collection of field data and our changing ideas concerning appropriate theory." This valuable experience shared by Walsham and Sahay (1999) highlights the importance of the 'open-minded principle' (no preconception) in GTM.

### 4.7.2. Comparative

The next level of literature engagement in the GTM study occurs, when based on the data analysis, it is found that the data obtained have relevancies with parts of the literature for comparisons between the two. This comparison is conducted to obtain more focused concepts that emerged from the field (Martin, 2006).

### 4.7.3. Integrative

During the mature stage of the theory development process, it is time to engage with the relevant literature at the theoretical level of comparisons to formulate the final theory.

### 4.7.4. Transcendent

The primary objective of literature engagement in this phase is to develop a formal theory that transcends the substantive theory (Martin, 2006). In this 'transcendent' phase, the researcher, guided by the data collected, engages with more literature beyond the substantive area being studied, such as psychology, sociology, and others.

In summary, GTM provides a set of principles and strategies for the data analysis process, which covers: (1) how to code (open and selective coding processes), (2) how to sample for more data (theoretical sampling strategy), which is guided by the emergence of concepts from data, (3) how to engage with the literature throughout the study, and (4) the whole process is facilitated by the use of memo writings.

## 5. THEORY BUILDING EXPERIENCE USING CASE STUDIES

Based on the above discussion on GTM data collection and analysis processes, this section discusses an example of a theory-building study using data collected from three (3) case studies in Timor-Leste. This example of GTM use is presented in two main stages: *first*, the data collection process, which involves how case studies are selected based on their data richness, their similarity, and their differences, and *second*, the data analysis process, which involves how the collected data are analyzed according to the GTM procedures presented in the previous section.

### 5.1. GTM using Case Study: Data Collection Process

During the early stage of the research, several potential cases were identified using publicly available information from the official websites and news outlets. In the end, based on theoretical sampling strategy, three case studies were selected in this study.

Table 1 summarizes the characteristics of the three case studies. Following the research ethics protocol of this study, all names of individuals, projects, and institutions are presented using pseudonyms.

| First Case: Alpha Project | Second Case: Beta Project | Third Case: Gamma Project |
|---|---|---|
| ● Implemented in the finance sector<br>● Funded by the host institution<br>● Proprietary system<br>● Supplied by a for-profit multinational corporation | ● Implemented in the justice sector<br>● Funded by a donor country<br>● Open-source system<br>● Supplied by a for-profit multinational corporation | ● Implemented in the education sector<br>● Co-funded by UN agencies (Delta and Epsilon)<br>● Open-source system<br>● Supplied by individual international consultants |
| **Table 1: Summary of Characteristics of the Three Case Studies** | | |

At first, the initial concepts identified in the preliminary literature review were used as 'a beginning foothold' (Glaser & Strauss, 1967, p. 45) for this research to facilitate the early identification and selection of the first case (the foundation case). A foundation case is a case selected from significant ICT4D projects, one that will provide the initial set of data for this study.

In this study, the first case was selected based on its potential data richness because the Alpha project has the following characteristics: (1) the E-gov technology was implemented in a very crucial area within the Government of Timor-Leste, the Finance Ministry which handles the State's annual budget execution process, (2) the project was funded by the host government, and (3) the E-gov technology was delivered by a for-profit multinational enterprise based in Canada.

Using the theoretical sampling strategy, the second case was selected based on findings that emerged from the first case; and then the final case was selected based on the combined findings of the previous cases. For each case, data were collected using interviews, observations, informal conversations, electronic correspondence, and relevant project documents.

### 5.2. GTM using Case Study: Data Analysis Process

The data analysis process in this research followed the double-back steps outlined in the earlier section. The following strategy and principles in GTM are used throughout the analysis process: theoretical sampling strategy, constant comparison, conceptual emergence, memo writing, theoretical sorting, and the use of literature throughout the study. These sets of steps interplay with each other throughout the three case studies.

#### 5.2.1. First Case Study: Alpha Project

The process of selecting this first case already involved the theoretical sampling strategy because the Alpha case was selected based on its potential data richness. The specific GTM data analysis process for this case is summarized below.

**Theoretical Sampling Strategy:** In this first case, the sampling strategy involved only one strategy: the 'intra-case sampling strategy', because the focus of data collection was to develop and saturate concepts and their properties within this case only (Fernández & Lehmann, 2011). It is within this case that the 'substantive coding' process of GTM took place. At first, it solely involved the 'open coding' process, where data were analyzed openly to allow possible concepts to emerge. Later on, when the main concepts and their properties (project phases and sub-processes) had emerged, data sampling and coding were conducted selectively based on the emergent concepts ('selective coding'). This process continued until data had repeated itself and data collection from different sources yielded no new concepts and/or properties. Memos were used to record understanding of data (alteration/confirmation of concepts).

**Constant Comparison Strategy**: The primary strategy used throughout the data analysis process was the comparison strategy. For each interview/observation transcript, the author openly conducted the analysis process by investigating the transcript line by line, sentence by sentence, or paragraph by paragraph, and so on, to develop concepts. Each new indicator that pointed to a specific concept was constantly compared with other indicators. At the same time, any new potential concept was also compared with existing concepts to determine their differences and/or similarities. This process allows to either: (1) confirm or revise an existing concept, (2) develop a new concept that fits the indicator, and (3) develop a higher level of concept if the new indicator pointed to the existence of a concept that is higher than the existing concepts. This coding process continued until the main pattern emerged and the core concepts were developed.

From this Alpha case, one pattern emerged from data: all project stakeholders are trying to deliver an E-gov technology wanted by the Minister. There were also three main project phases identified together with their sub-processes in each of the project phases.

The pattern and core concepts and their properties were constantly compared to new slices of data from this case until data repeated itself and yielded no new higher abstract concepts. This means that the existing core concepts became saturated.

**The use of literature in the first case**: Literature was also used in this Alpha case with the primary objective of comparing the emergent concepts with the previous findings in the literature; this step is called the comparative stage (Martin, 2006; Urquhart & Fernández, 2013). For instance, the concept of forms of capital by Bourdieu (1986) emerged as a valuable input to further fine-tune the analysis process when it was becoming clear that all of the actors involved had an interest in the forms of capital.

When saturation has been reached in the intra-case sampling stage, it is time to bring in more 'slices of data' from other cases (the 'inter-case sampling' strategy). The next case should have maximum differences compared to the first case to increase the potential for the emergence of new concepts and/or properties of concepts. It has to be noted that these "maximum differences" between projects must be within the focus (scope) of the research, which is the E-government projects in Timor-Leste. For this reason, the Beta case was selected.

### 5.2.2. Second Case Study: Beta Project

A foreign donor country funded the Beta project, and the system was delivered to institutions in the justice sector in Timor-Leste. This E-gov technology was delivered by a multinational open-source enterprise based in the United States of America and India. The specific GTM data analysis process for this case is summarized below.

**Theoretical Sampling Strategy:** In this second case, the sampling strategy involved mainly the 'inter-case sampling strategy' because the focus of data collection was to confirm, reject or revise the concepts and their properties that emerged from the first case (Fernández & Lehmann, 2011). Within this case, the primary coding process was still the 'substantive coding' and involved mainly the 'selective coding' process because the coding was based on findings from the first case.

A new pattern emerged from this Beta project implementation where all the stakeholders are trying to deliver a sustainable E-gov technology to the institutions in the justice sector. However, data from the Beta case confirm that the three main project phases emerged from the Alpha case. For this research project, the second pattern was selected as the primary focus simply because this finding fits with the overall concern in the ICT4D research on the sustainability issue of the ICT4D initiatives. This process continued until data had repeated itself, and further data collection from different sources, in this case, yielded no further concepts and/or their properties. The subsequent case, the third case, was selected to confirm and enrich the findings that emerged from the previous two case studies.

**Constant Comparison Strategy:** In the second case, the constant comparison was also the primary strategy used. The coding process continued until the core concepts, and their properties were developed. The pattern and core concepts and their properties were constantly compared to new slices of data from this case until data repeated itself (saturation of data).

**The use of literature in the second case:** In the second case, literature is also used with the primary objective of comparing the emergent concepts: the comparative stage (Martin, 2006; Urquhart & Fernández, 2013). In this case study, the author began to engage with the literature on sustainability issues in ICT4D research. When saturation has been reached in this first inter-case sampling strategy because no more new concepts or properties of concepts emerged from data collected from this second case, it was time to bring in more 'slices of data' from other cases (Fernández & Lehmann, 2011). Similar to the first inter-case sampling strategy for the second case, the next case should also have maximum differences compared to the first and second cases to increase the potential for the emergence of new concepts or properties of concepts. For this reason, the Gamma case was selected.

### 5.2.3. Third Case Study: Gamma Project

The Gamma case was also selected using the theoretical sampling strategy, based on its maximum differences compared to the first and second cases. The Gamma case was co-funded by two United Nations (UN) agencies operating in Timor-Leste, and the system was delivered to the Timorese

Education Ministry. Different from the two previous cases, the E-gov technology in the Gamma case was delivered by individual international consultants. The specific GTM data analysis process for this case is summarized below.

**Theoretical Sampling Strategy:** Same with the second case, the sampling strategy in this third case involved mainly the 'inter-case sampling strategy' to confirm, reject or revise the concepts and their properties that emerged from the first and the second cases (Fernández & Lehmann, 2011). Within this case, the primary coding process was still 'substantive coding' and involved mainly the 'selective coding' process based on findings from the first two cases. This process continued until data had repeated itself and yielded no further results that may be sufficient to confirm, revise or reject the existing concepts (or their properties). At the end of this case study, the analysis in this research project was based mainly on the 'theoretical coding' process to develop relationships between concepts that emerged from the three cases.

**Constant Comparison Strategy:** In the third case, the constant comparison strategy was also used. This coding process took place by constantly comparing the existing pattern in the Gamma case and the core concepts that emerged from the second case until data repeated itself and yielded no new higher abstract concepts. This means that the existing core concepts have become saturated. In this Gamma case, another project pattern emerged where all the stakeholders are trying to deliver the E-gov technology sponsored by the UN agencies. However, there were no further significant findings emerged to alter or enrich the findings developed previously.

**The Use of Literature in the Third Case:** At the end of this third case, the existing core concepts that were developed in the first and second cases had been saturated. During this mature stage of the theory development process, it was time to engage with the relevant literature at the theoretical level of comparisons to formulate the final theory (Urquhart & Fernández, 2013).

This study developed a substantive theory of "cultivating sustainability" of E-gov projects in developing countries based on the three case studies. This theory explains how an ICT4D project, specifically an E-gov project, can be designed in a way to achieve its sustainability throughout the project implementation process.

## 6.    CONCLUSION

This paper outlines the concepts around GTM and showcases a theory development process using data collected from three E-gov case studies in Timor-Leste. There have been a few theory development studies conducted in the ICT4D area (Avgerou, 2008; Heeks, 2006; Sahay & Walsham, 1995). Therefore, the theory development process presented in this paper serves as a reference for research students and early career academics in the ICT4D area to explore potential research methodologies for their research. In the end, it may help to encourage new studies to develop more native and indigenous theories of ICT4D. Thus, allowing us to share the pearls of wisdom of societies in developing countries with the rest of the world.